\documentclass{ws-procs9x6}

\newcommand{\refeq}[1]{(\ref{#1})}

\begin{document}

\title{Lorentz Violation and Topological Defects}

\author{Michael D. Seifert}

\address{Department of Physics, Williams College, Williamstown, MA
  01267, USA \\
  E-mail: ms10@williams.edu}

\begin{abstract}
  If Lorentz symmetry is broken, it must have occurred dynamically,
  via a vector or tensor field whose potential energy forces it to
  take on a non-zero background expectation value ``in vacuum''.  If
  the set of minima of this potential (the \emph{vacuum manifold}) has
  a non-trivial topology, then there can arise \emph{topological
    defects}: stable solutions in which the field approaches different
  potential minima as we go to infinity in different directions.  I
  discuss the current status of research into these topological
  defects in the context of Lorentz symmetry breaking, including
  recent results concerning the birefringent light-bending of monopole
  solutions, and the search for models supporting cosmic-string and
  domain-wall defects.
\end{abstract}

\bodymatter

It has been known for some time now \cite{AKnogo} that if Lorentz
symmetry is to be broken and spacetime is to be well-described by
Riemannian geometry, then the breaking of Lorentz symmetry must be
spontaneous.  The Lagrangian for a model in which this occurs will
have the general form
\begin{equation}
  S = \int d^4 x \sqrt{-g} \, \left[ \frac{R}{16 \pi G} - (\nabla \mathcal{T}) (\nabla
    \mathcal{T}) - V(\mathcal{T}) \right],
  \label{mds:lvlag}
\end{equation}
where $R$ is the Ricci scalar and $\mathcal{T}$ is a tensor field
(whose indices have been suppressed for the sake of generality).  If
the potential term $V(\mathcal{T})$ is constructed in such a way that
\begin{equation}
  V(\bar{\mathcal{T}}) = \frac{\delta  V(\bar{\mathcal{T}})}{\delta
    g^{ab}} = \frac{\delta  V(\bar{\mathcal{T}})}{\delta
    \mathcal{T}} = 0
  \label{mds:vacmanconds}
\end{equation}
for some non-zero tensor $\bar{\mathcal{T}}$, then a solution of the
equations of motion is $g_{ab} = \eta_{ab}$ and $\mathcal{T} =
\bar{\mathcal{T}}$ everywhere in spacetime.  In other words, we have
Minkowski spacetime, but with an additional geometric structure (due
to the constant but non-zero tensor field $\mathcal{T}$) upon which we
can ``hang'' Lorentz-violating effects.

Since the Lagrangian \refeq{mds:lvlag} is Lorentz- and
diffeomorphism-invariant, the background value $\bar{\mathcal{T}}$ of
the tensor field in such a theory will not be unique.  Rather, there
will be some set of possible tensor values that will satisfy the
conditions \refeq{mds:vacmanconds}.  This set of tensor values forms a
manifold in field space, called the \emph{vacuum manifold}.  Its
dimension and shape will depend on the precise form of the potential.
As an example, suppose that $\mathcal{T}$ is a vector field $A^a$, and
that $V(A^a) = \lambda (A^a A_a - \mathcal{C})^2/4$.  The vacuum
manifold will then be a three-dimensional hyperboloid in the
four-dimensional field space, consisting of all vectors of a given
norm $A^a A_a = \mathcal{C}$.  If $\mathcal{C} >0$, then all these
vectors will be spacelike and the hyperboloid will be homeomorphic to
$S^2 \times \mathbb{R}$; if $\mathcal{C} < 0$, then the hyperboloid
will consist of two disconnected hypersurfaces, each homeomorphic to
$\mathbb{R}^3$ (i.e., the past-directed and future-directed vectors.)

If the topology of the vacuum manifold is non-trivial, it is possible
that static \emph{topological defect} solutions can
arise. \cite{VilShel} These solutions have the property that the field
takes on different vacuum values at different ``points at infinity'',
but that the field configuration at infinity cannot be smoothly
extended to all of space without the field leaving the vacuum
manifold.  There results a configuration which is (at least)
metastable, but whose stress-energy tensor is non-zero.  The type of
solution that arises will depend on the topology of the vacuum
manifold; if there is a non-contractible $S^n$ in the vacuum manifold
(with $n \leq 2$), localized topological defect solutions are
possible.  A non-contractible $S^0$ can give rise to \emph{domain
  walls} separating two regions with different vacuum field values; a
non-contractible $S^1$ can lead to filament-like \emph{cosmic
  strings}; and a non-contractible $S^2$ can support point-like
\emph{monopoles}.  (In more precise mathematical terms, these cases
correspond to the vacuum manifold having non-vanishing homotopy groups
$\pi_0$, $\pi_1$, or $\pi_2$ respectively.)

The question then arises of whether the vacuum manifold for a
Lorentz-violating tensor field $\mathcal{T}$ could have the topology
necessary to support a topological defect.  This question was answered
in the affirmative in my previous work.\cite{MDStopdef} If we assume
that the vacuum manifold is of the form $\mathcal{T}_{abc\dots}
\mathcal{T}^{abc\dots} = \mathcal{C}$, only three types of irreducible
tensor fields have the right topology to support topological defects:
vectors, antisymmetric two-tensors, and symmetric trace-free
two-tensors.

Of these types of fields, topological defects have been found for one
of them: an antisymmetric tensor field $B_{ab}$ with a flat spacetime
Lagrangian of the form
\begin{equation}
  \mathcal{L} = - \frac{1}{12} F_{abc} F^{abc} - \frac{\lambda}{2} (B^{ab}
  B_{ab} - b^2)^2,
  \label{mds:astensoraction}
\end{equation}
where $F_{abc} \equiv 3 \partial_{[a} B_{bc]}$.  A spherically
symmetric static monopole solution exists in this theory, and its
properties have been studied.\cite{ASmonopole} In this solution,
$B_{\theta \phi}$ interpolates smoothly from zero at $r = 0$ to the
vacuum manifold as $r \to \infty$; all other components of $B_{ab}$
vanish.  We can also ``turn on'' gravity by including the Ricci scalar
in the above action.  The spacetime curvature which results due to the
stress-energy tensor of $B_{ab}$ would then be detectible due to its
influence on light rays.  It can be shown that the effects of
gravitational redshift would be within a couple of orders of $\epsilon
\equiv 4 \pi G b^2$, and that angular deflection $\delta \phi$ of
light rays is constant to leading order in the impact
parameter.\cite{ASmonopole}

More recently,\cite{KLMDS} Kamuela Lau and I have been investigating
the behavior of light rays travelling near a monopole when there is a
direct coupling between the Maxwell field and the monopole field
$B_{ab}$:
\begin{equation}
  \mathcal{L} \supset - \frac{1}{4} F_{ab} F^{ab} - (k_F)^{abcd}
  F_{ab} F_{cd}
\end{equation}
where $F_{ab}$ is the Maxwell field strength,
\begin{equation}
(k_F)^{abcd} = \xi \left( B^{ab} B^{cd} - \frac{1}{12} (\eta^{ac}
\eta^{bd} - \eta^{ad} \eta^{bc}) B^{ef} B_{ef} - B^{[ab} B^{cd]}\right),
\end{equation}
and $\xi$ is a coupling constant.  The modified Maxwell's equations
that arise are equivalent to Maxwell's equations in an inhomogeneous
birefringent medium.\cite{MMAKMaxwell} We can then use a
geometric-optics approximation to study the paths of light rays.  The
result is that a distant source will be split into two images in the
presence of a monopole solution, each image being entirely polarized.
One polarization will travel through the monopole in a straight line;
the other will be bent, and so will appear at a different location on
the sky.  The angle between these two images will generically be of
order $\xi b^2$, where $b$ is the vacuum magnitude of the field
$B_{ab}$.

Another student, Brandon Ling, has been conducting a more general
survey of the possible types of vacuum manifolds and topological
defects that might be found.  Specifically, he has examined the
topology of the vacuum manifolds for an antisymmetric tensor field
$B_{ab}$ with a quartic potential of the form
\begin{equation}
  V(B_{ab}) = \alpha X^2 + \beta XY + \gamma Y^2 + \delta X + \lambda
  Y, 
  \label{mds:genASpot}
\end{equation}
where $X = B_{ab} B^{ab}$ and $Y = B_{ab} B_{cd} \epsilon^{abcd}$.
(Note that the potential in \eqref{mds:astensoraction} corresponds to
the case $\beta = \gamma = \lambda = 0$, up to an additive constant.)
He has found that the only non-contractible spheres in these vacuum
manifolds are two-spheres; no non-contractible circles can exist, and
the vacuum manifolds are always connected.  This implies that
antisymmetric tensors cannot give rise to cosmic strings or domain
walls.  Remarkably, it can be shown that in most cases the vacuum
manifold for the potential \eqref{mds:genASpot} is homeomorphic to $TS^2$,
the tangent bundle on the two-sphere.  We are now exploring which
parameter values in this allow for monopole solutions,
and what the properties of these solutions will be.

\section*{Acknowledgments}
I would like to thank Williams College for their financial support in
attending this conference.


\begin{thebibliography}{9}

\bibitem{AKnogo}  V.\ A.\  Kosteleck\'y, Phys.\ Rev.\ D {\bf 69},
  105009 (2004).

\bibitem{VilShel} A.\ Vilenkin and E.\ P.\ S.\ Shellard, {\it Cosmic
    Strings and Other Topological Defects}, Cambridge University
  Press, Cambridge, 1994.

\bibitem{MDStopdef}  M.\ Seifert, Phys.\ Rev.\ D {\bf 82}, 125015 (2010).

\bibitem{ASmonopole} M.\ Seifert, Phys.\ Rev.\ Lett.\ {\bf 105},
  201601 (2010);  X.\ Li, P.\ Xi, and Q.\ Zhang, Phys.\ Rev.\ D {\bf
    85} 085030 (2012).

\bibitem{KLMDS} K.\ Lau and M.\ Seifert.  In preparation.

\bibitem{MMAKMaxwell} V.\ A.\  Kosteleck\'y and M.\ Mewes, Phys.\
  Rev.\ D {\bf 66}, 056005 (2002).

\end{thebibliography}
\end{document}